\documentclass[twocolumn,secnumarabic,amssymb, nobibnotes,superscriptaddress,showpacs, prb]{revtex4-1}
\usepackage{bm}%
\usepackage{color}
\usepackage{graphicx}
\usepackage{subfigure}

\begin{document}
\title{First principles study of the influence of (110)-oriented strain on the ferroelectric properties of rutile TiO$_2$}%
\author{A. Gr\"unebohm}
\email[Electronic mail: ]{anna@thp.uni-duisburg.de}
\affiliation{Faculty of Physics and Center for Nanointegration, CeNIDE, University of Duisburg-Essen, 47048 Duisburg, Germany}
\author{C. Ederer}
\affiliation{School of Physics, Trinity College, Dublin 2, Ireland}
\author{P. Entel}
\affiliation{Faculty of Physics and Center for Nanointegration, CeNIDE, University of Duisburg-Essen, 47048 Duisburg, Germany}

\date{June, 2011}
\begin{abstract}
We use first principles density functional theory to investigate the
softening of polar phonon modes in rutile TiO$_2$ under tensile
(110)-oriented strain. We show that the system becomes unstable
against a ferroelectric distortion with polarization along (110) for
experimentally accessible strain values. The resulting polarization,
estimated from the Born effective charges, even exceeds the bulk
polarization of BaTiO$_3$. Our calculations demonstrate the different
strain dependence of polar modes polarized along (110) and (001)
directions, and we discuss the possibility of strain engineering the
polarization direction, and the resulting dielectric and piezoelectric
response, in thin films of TiO$_2$ grown on suitable substrates.
\end{abstract}
\maketitle


Rutile TiO$_2$ is an {\it{incipient}} ferroelectric
material,\cite{Lee,Traylor} and it has been shown via {\it{ab initio}}
simulations that a ferroelectric transition can be induced by either
negative pressure or (001)-oriented
strain.\cite{Montanari,Montanari2,Liu,felich} An experimental
realization of ferroelectric TiO$_2$ would have great technological
impact, since this material is cheap and optimized processing
techniques are readily available. TiO$_2$ is currently used for a
large variety of applications such as solar cells or pigments. Besides, due to its high dielectric constant ($\epsilon
= 251$ at low temperatures \cite{Samara}), TiO$_2$ is also used in
dielectric devices. Hence, the possibility to tune the electric
permittivity with or without inducing a ferroelectric transition would
be of great technological importance.

Most applications are based on the (110) surface, which can be strained in either ($\bar{1}$10) or (001) direction by
growth on a suitable substrate.
Until now, only the influence of isotropic and (001)-oriented strain
on a ferroelectric state with polarization along (001) has been
investigated for bulk rutile.\cite{Montanari,felich,Mitev,Liu} The
resulting ferroelectricity can be attributed to the only polar phonon
mode polarized in (001) direction, $A_{2u}$, which softens if the Ti-O
distances are enlarged. Interestingly, it was predicted recently that
the $A_{2u}$ mode is insensitive to (110) strain whereas an acoustic
phonon mode softens under these strain conditions.\cite{Mitev} Apart
from $A_{2u}$, rutile possesses additional low energy phonon modes
which are polarized along (110), and for which a detailed
understanding is still missing.

Here, we investigate the role of (110)-oriented strain on the
ferroelectric properties of rutile TiO$_2$. The main goal of our work
is a systematic investigation of the influence of (110) strain on
polar modes in (001) and (110) direction. We show that the
paraelectric state of rutile is destabilized under tensile strain and
a polarization along (110) is induced. Furthermore, a polarization in
(001) direction can be obtained if the ionic shifts along the (110)
direction are suppressed.

\begin{figure*}
\includegraphics[width=0.65\textwidth]{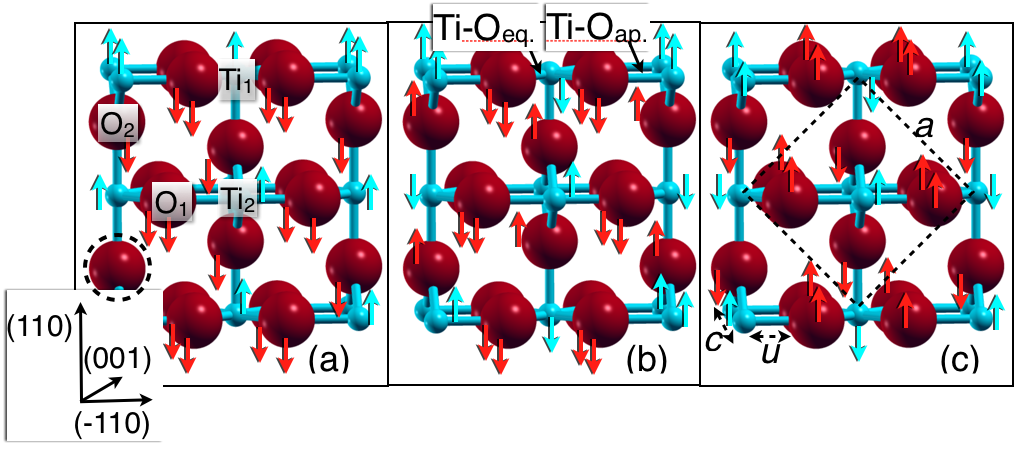} 
\caption{Atomic structure of bulk rutile in the
  $\sqrt{2}\times\sqrt{2}\times1$ supercell. Solid arrows indicate the
  atomic shifts for the three (110) polarized $B_{3u}$-modes used in
  our analysis, sorted in ascending order of their energies. (a)
  $B_{3u,1}$ mode, indices mark inequivalent Ti/O sublattices under
  (110) strain, and the dotted circle marks the O atom which has been
  shifted previous to the structural optimization; (b) $B_{3u,2}$
  mode; (c) $B_{3u,3}$ mode. Dotted lines and arrows mark the
  primitive unit cell, the lattice constants $a$ and $c$, and the
  internal structural parameter $u$.}
\label{fig:eumoden}
\end{figure*}

All results presented here are obtained from self-consistent density
functional theory calculations employing the ``Vienna Ab Initio
Simulation Package'' (VASP) \cite{Kresse1} with a plane wave basis,
projector augmented wave potentials,\cite{Blochl} and the generalized
gradient approximation (GGA) in the formulation of Perdew, Burke and
Ernzerhof.\cite{PBE} For Ti (O) atoms the $4s3d$ ($2s 2p$) electrons
are treated as valence. Ionic positions are optimized until all forces
are smaller than 0.01~eV/{\AA} and a $\sqrt{2} \times \sqrt{2} \times 1$
supercell containing 4 TiO$_2$ units, see Fig.~\ref{fig:eumoden}, is
used. Born effective charges have
been calculated using density functional perturbation
theory.\cite{bornstoer}


The rutile crystal structure of TiO$_2$ is tetragonal with space group
symmetry $P4_2/mnm$ (see the primitive unit cell inserted in
Fig.~\ref{fig:eumoden}c).  Each Ti-atom is 6-fold coordinated by
O-atoms, with 4 short {\it{equatorial}} Ti-O bonds and two longer
{\it{apical}} bonds, see Fig.~\ref{fig:eumoden}b. The exact Ti-O
distances are determined by the lattice constant $a$, the tetragonal
$c/a$ ratio, and the internal structural parameter $u$. From our relaxations we obtain
$a=4.6640$~{\AA} (4.587~{\AA}), $c/a=0.6366$ (0.644), and $u=0.3047$
(0.305). For comparison, experimental results from
Ref.~\onlinecite{Burdett}, using neutron diffraction at 15~K, are
given in brackets.

As mentioned in the introduction, one polar optical $A_{2u}$ mode with
polarization along (001) exists in rutile. The corresponding
ferroelectric state can be stabilized by increasing the equatorial
Ti-O bond length, since this reduces the short range repulsion between
the two ions.\cite{Montanari,Montanari2,Liu}
Furthermore, there are three pairs of doubly-degenerate polar optical
phonon modes of $E_u$ symmetry, which correspond to atomic
displacements within the $(110)$-$(\bar{1}10)$ plane.\cite{Lee} In
Ref.~\onlinecite{Montanari} it has been shown that a uniform lattice
expansion causes a softening of the energetically lowest $E_u$ mode,
although this effect is smaller than the corresponding change for the
A$_{2u}$ mode.

The application of (110) strain reduces the symmetry to the
orthorhombic space group $Cmmm$, with two inequivalent Ti (Ti$_1$,Ti$_2$) and O (O$_1$, O$_2$) sublattices. The imposed strain
changes the {\it{apical}} Ti$_1$-O$_2$ and the {\it{equatorial}}
Ti$_2$-O$_2$ bond distance, whereas the {\it{equatorial}} Ti$_1$-O$_1$
and the {\it{apical}} Ti$_2$-O$_1$ bonds are not modified, see
Fig.~\ref{fig:eumoden}(a).  As a result, the twofold degeneracies among
the six optical $E_u$ modes are lifted and they split into three
$B_{2u}$ and three $B_{3u}$ modes with atomic shifts along
($\bar{1}10$) and (110), respectively.

For further analysis of the atomic displacements it is convenient to
decompose the $3\times N$ dimensional displacement vector ${\bf{Y}}$
into a linear combination of three mode patterns ${\bf{B}}_{3u,i}$ and
an additional relaxation ${\bf{R}}$,
 \begin{equation}
{\bf{Y}}=A\cdot {\bf{B}}_{3u,1} + B\cdot {\bf{B}}_{3u,2} + C\cdot {\bf{B}}_{3u,3} +{\bf{R}}.
\label{eq:mode}
\end{equation}
Here, $N$ is the number of atoms within the supercell. We use a set of
$B_{3u,i}$ modes with displacement patterns as sketched in
Fig.~\ref{fig:eumoden} and the same magnitude of displacements on all
sublattices. These three modes, together with the corresponding
acoustic mode, form a complete basis for representing the $B_{3u}$
eigenmodes of the strained rutile structure. Since we are only
interested in the three optical modes, we do not consider the acoustic
mode in our analysis. We note that even though the displacement
amplitudes of the various sublattices in the exact phonon eigenvectors
cannot be fully determined by symmetry arguments alone, the three mode
vectors depicted on Fig.~\ref{fig:eumoden} are indeed relatively close
to the actual $E_u$ eigenvectors of unstrained TiO$_2$.

\begin{figure}[h]
\includegraphics[width=0.35\textwidth]{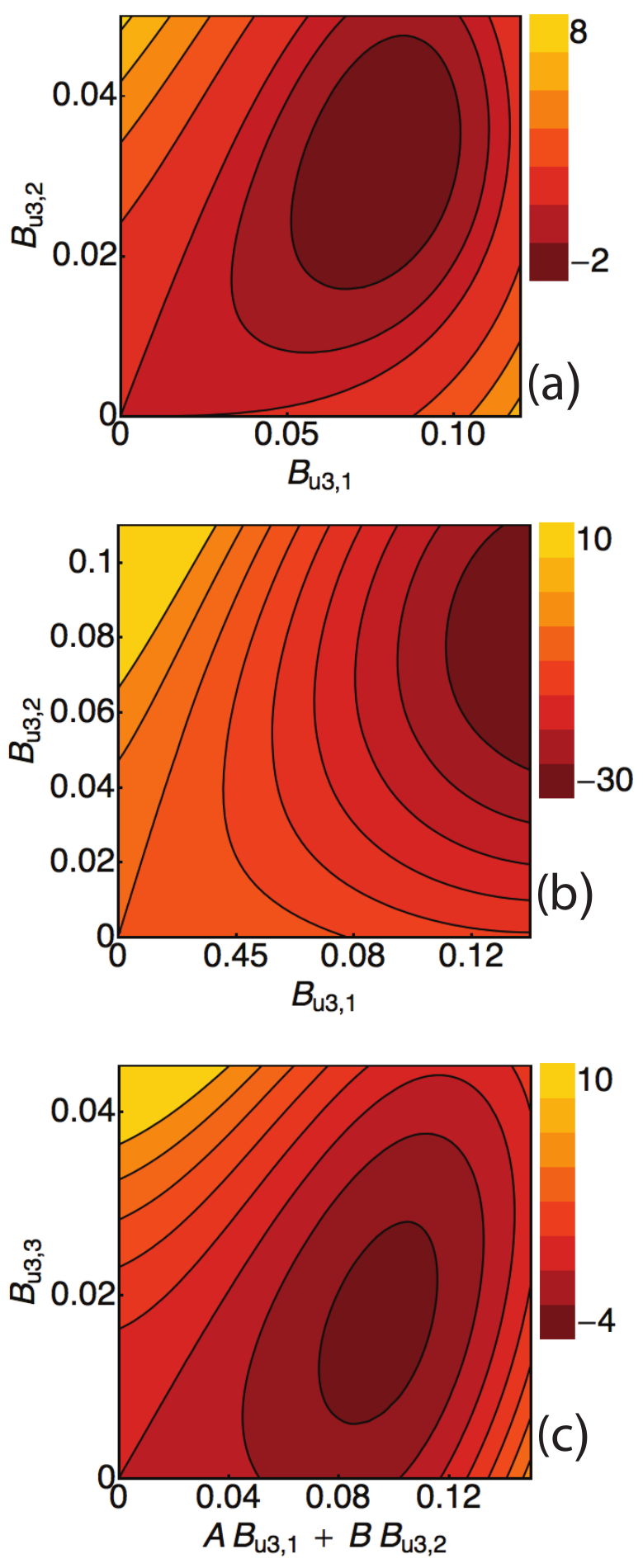}
\caption{Energy landscapes for static displacements along the
  $B_{3u,i}$ mode vectors corresponding to 2\,\% strain (a),(c) and 5\,\% strain (b).
  Amplitudes of the modes are given in {\AA} and energies in meV/TiO$_2$ unit relative to the unrelaxed strained configuration.}
\label{fig:e110bulk}
\end{figure}

To explore the energy landscape of (110)-strained TiO$_2$ as function
of the $B_{3u}$ optical modes, we perform total energy calculations with fixed atomic positions, and include different
amounts of the three distortion modes $B_{3u,i}$. The system is softest
along $B_{3u,1}$, but a combination of $B_{3u,1}$ and $B_{3u,2}$ is
required to produce a stable ferroelectric energy minimum, see
Fig.~\ref{fig:e110bulk}.
This energy minimum is shifted to larger amplitudes for increasing
strain (compare Figs.~\ref{fig:e110bulk}a and \ref{fig:e110bulk}b),
but no major differences appear for higher strain values.
In the following we focus our discussion on the case of 2\,\% strain.

In order to locate an approximate energy minimum, we first identify
the most favorable $B_{3u,2}$ amplitude as function of $B_{3u,1}$, and
then vary the amplitude of $B_{3u,3}$ for the resulting optimal linear
combinations of $B_{3u,1}$ and $B_{3u,2}$. The so-obtained energy
landscape is shown in Fig.~\ref{fig:e110bulk}c. From this we derive an
approximate energy minimum at ${\bf{Y}}_0$ corresponding to
$A=0.096$\,{\AA}, $B=0.035$\,{\AA}, $C=0.017$\,{\AA}, and
${\bf{R}}=0$.\\
The results of our full structural relaxations for different
magnitudes of (110) strain are listed in Table~\ref{tab:110}. 
The symmetry of the system has been reduced to $Pmm2$ prior to the relaxation by shifting one of the O$_2$
atoms 0.002\,{\AA} along (110), see Fig.~\ref{fig:eumoden}(a).

The atomic relaxations confirm that for tensile (110) strain the
paraelectric state is destabilized by a polar shift of the sublattices
against each other in (110) direction. This shift, which is well
approximated by ${\bf{Y}}_0$ (see Table~\ref{tab:110}), leads to
alternating expansion/contraction of the short Ti$_2$-O$_2$
bonds.
However, since a further increase of this bond length (besides the
increase resulting from the overall (110) strain) is unfavorable,
additional atomic displacements corresponding to an $M$ point zone
boundary mode lead to different (110) shifts of the two O$_2$
atoms. Furthermore, a small relaxation of the O$_1$ atoms in
($\bar{1}10$) direction (not included in Table~\ref{tab:110}) is also
superimposed to ${\bf{Y}}_0$. As a result, the Ti$_1$-O$_1$ bond
length is approximately conserved, even though these ions are shifted
against each other in (110) direction. This lowers the energy for the
(110) shift and thus the amplitudes of the (110) shift for these
sublattice are slightly increased compared to ${\bf{Y}}_0$.
\begin{table}
\caption{Amplitudes $x$ (in {\AA}) of atomic shifts in (110) direction
  under tensile strain. The energy differences $\Delta{E}$ relative to
  the paraelectric state at the same strain are given in meV/TiO$_2$ unit, and the resulting polarization $P_s$ in $\mu$C/cm$^{-1}$.}
\label{tab:110}
\begin{tabular}{lcccccc}
\hline
\hline
Strain&$x(\text{Ti}_1)$&$x(\text{Ti}_2)$&$x(\text{O}_1)$&$x(\text{O}_2)$&$\Delta E$ &$P_s$ \\
\hline
1.02\footnote{Shift along ${\bf{Y}}_0$ without further atomic relaxation.}& 0.09&	0.04 &    $-$0.02&$-$0.04   &	4.8 	  	&	  33 \\
1.02	& 0.10&	0.04 &    $-$0.03& $-$0.03/$-$0.05   &	6.0   	  	&	  35 \\
1.03	& 0.13&	0.05   &  $-$0.03&    $-$0.04/$-$0.08  &	15.2		&	  46\\
1.04	& 0.16&	0.05    & $-$0.03&      $-$0.04/$-$0.10 &  	29.8	&		  54\\
1.05	& 0.19&	0.05    & $-$0.03&     $-$0.05/$-$0.13   & 	49.3	&	  	  62\\
\hline
\hline
\end{tabular}
\end{table}
\begin{figure*}
\includegraphics[width=0.75\textwidth]{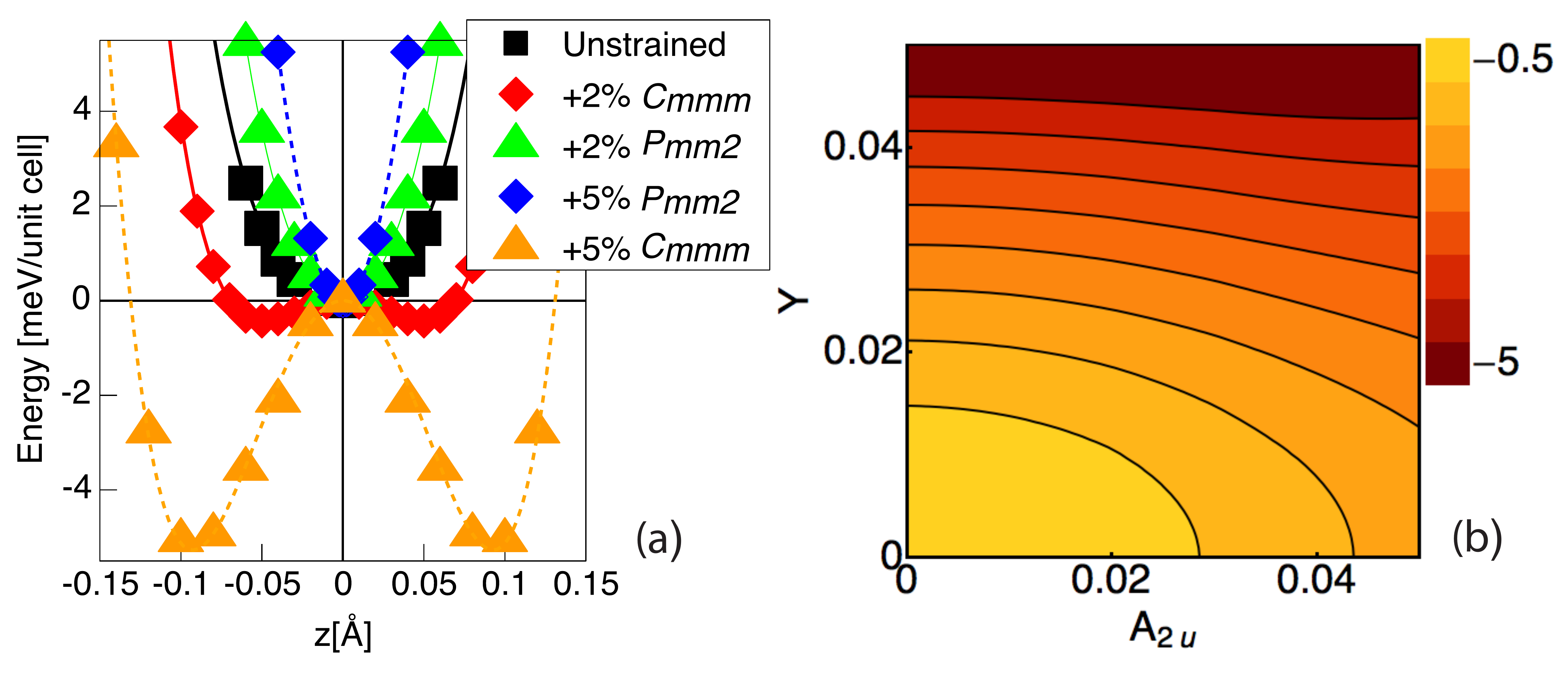}
\caption{(a) Total energy change for static atomic displacements along
  the $A_{2u}$-mode under uniaxial (110) strain for atomic positions relaxed with imposed  $Cmmm$/$Pmm2$ symmetry. Here, only the second symmetry class allows for polarization in (110) direction.  (b) Energy landscape for
  mixed displacements along ${\bf{Y}}_0$ and $A_{2u}$ for 2~\% tensile
  strain in (110) direction.}
\label{fig:e110bulk2}
\end{figure*}

The resulting spontaneous polarization can be
estimated by
\begin{equation}
P_{(110)}=\frac{1}{V}\sum_i x(i)\cdot Z^{*}_{(110),i}.
\label{eq:pol}
\end{equation}
Here, the summation is over all atoms in the unit cell with volume
$V$, $x(i)$ represents the shift of atom $i$ in (110) direction, and
$Z^{*}_{(110),i}$ the corresponding principal value of the Born
effective charge tensor for undistorted bulk. As the Born charges in TiO$_2$ are
quite stable with respect to small modifications of the Ti-O
distances,\cite{felich} this approximation is sufficient for a
qualitative discussion. We obtain $Z^{*}_{(110)} = 7.59|e|$ ($-5.14
|e|$) for Ti$_1$ (O$_2$), which has the apical bond aligned along
(110), whereas we obtain $Z^{*}_{(110)} = 5.35|e|$ ($-1.32|e|$) for
Ti$_2$ (O$_1$)
which is in qualitative agreement to $Z^{*}_{(110)}=7.34|e|$ ($-4.98|e|$) along the {\it{aptical}} bond, respectively $Z^{*}_{(110)}=5.34|e|$ ($-1.36|e|$) for the perpendicular direction obtained by Lee {\it{et al.}}.\cite{Lee}.\\
With Eq.~\ref{eq:pol} we calculate a sizeable polarization,
which is already in the technical relevant range for a strain of 2~\%,
see Table~\ref{tab:110}. The resulting ferroelectric state is 6 meV/formula unit lower in energy than the paraelectric state.
If one assumes that the ferroelectricity is related to exactly one degree of freedom at the $\Gamma$ point this correspond to 
a thermal energy of about 139~K. However, since the
depth of the energy well increases strongly with the imposed strain, a ferroelectric phase should be
experimentally accessible for about 5~\% tensile strain as our rough estimation already results in a thermal energy of 1146~K in this case.

For compressive (110) strain, the reduction of the Ti-O distances
induces a large short range repulsion, and a ferroelectric distortion
is unfavorable.  Instead, O$_2$/O$_1$ atoms increase the (110)/(-110)
component of their {\it{equatorial}} bond, due to the larger
compressibility of the {\it{apical}} bond. This is in agreement with
the hardening of the ferroelectric A$_{2u}$ mode under
compression.\cite{Montanari, felich}

We note that we could not reproduce the destabilization of the system
along an acoustic phonon mode under (110) strain found in
Ref.~\onlinecite{Mitev}, for which an energy gain of several meV has
been predicted, even if we use a $2\cdot\sqrt{2}\times2\cdot\sqrt{2}\times4$ supercell which is  commensurable with the displacement pattern of this mode.
 Possible explanations for this discrepancy are the
different potentials used and the resulting difference in lattice
parameters.  

Finally, we discuss the relationship between the polar modes with
polarization along (110) and (001), respectively. Since the
{\it{equatorial}} Ti-O bond length increases under tensile (110)
strain, the short range repulsion between these ions is reduced, and a
polar shift along (001) should become likely. Nevertheless, no
softening of the $A_{2u}$-mode under (110) strain has been observed in
previous {\it ab initio} investigations.\cite{Mitev} In contrast, our
calculations show that the polar mode in (001) direction can be
stabilized if the displacements along the $B_{3u}$ modes are
suppressed, see Fig.~\ref{fig:e110bulk2}.\cite{footnote} If the ions are not allowed to displace
along (110), a characteristic double well potential appears for
tensile (110) strain, and the well depth increases with increasing
strain. However, a coupling between soft modes corresponding to (110)
and (001) polarization exists, similar to perovskites such as
e.g. SrTiO$_3$.\cite{Antons} As a result, the ferroelectric mode
polarized in (110) direction disables polar distortions in (001)
direction. If the strain increases from 2\,\% to 5\,\%, and the
relative ionic shifts in (110) direction increase, more energy is
needed to shift the sublattices relative to each other along (001).
Fig.~\ref{fig:e110bulk2}b shows the energy landscape as function of
polar shifts along (001), $A_{2u}$, and along the optimized mode
vector ${\bf{Y}}_0$ in (110) direction.

While for pure (110)-oriented strain the energy gain corresponding to
the $A_{2u}$-mode is one order of magnitude smaller than the energy
gain corresponding to $B_{3u}$, a further softening occurs under
tensile (001) strain,\cite{Montanari} which is stronger for $A_{2u}$
than for $B_{3u}$. The exact location and depth of the energy minimum
within the ``$A_{2u}$-$B_{3u}$ plane'' can therefore be adjusted by
straining in both (110) and (001) directions, which can be achieved in
thin films by choosing an appropriate substrate. This opens up the
exciting possibility of ``strain engineering'' the polarization
direction in TiO$_2$, and the resulting dielectric and piezoelectric
response, similar to the case of perovskite
ferroelectrics.\cite{Antons,Dieguez/Rabe/Vanderbilt:2005}

In summary we have shown that a ferroelectric state can be stabilized
in rutile under tensile (110)-oriented strain. The main contribution
of the displacement pattern can be attributed to the $E_{u}$ phonon
modes of undistorted rutile. For 5\,\% strain a large polarization of
63\,$\mu$C/cm$^{-1}$ emerges in (110) direction. The depth of the
corresponding energy well is large in comparison to thermal
fluctuations at low temperatures, which suggests that the
ferroelectric state can indeed be observed experimentally. We note
that while the well-known small overestimation of unit cell volumes
within GGA may also slightly overestimate the tendency towards
ferroelectric distortions, qualitative trends are nevertheless
described properly and agree well with other
calculations.\cite{felich,Montanari2,Shojaee}

Furthermore, we have demonstrated that a ferroelectric state in (001)
direction can also be stabilized under tensile (110) strain, if the
polar shifts along the (110) direction are suppressed. This indicates
a coupling between optical modes polarized in (001) and (110)
directions, which allows to shift the relative energies of
ferroelectric states with polarization in (110) and (001) directions
under different strain conditions. Future investigation are necessary
in order to investigate the influence of additional (001) strain on
the ferroelectric modes in more detail. In addition, the effect of
surface-induced atomic relaxations at the real (110) surface on the
polar phonon modes is important and will be addressed in an upcoming
investigation.\cite{felich}

This work was supported by the Deutsche Forschungsgemeinschaft (SFB
445) and by Science Foundation Ireland (SFI-07/YI2/I1051).

\bibliographystyle{lit}
\bibliography{../Fe_rutil110/anna,claude}

\begin{thebibliography}{10}%
\makeatletter
\providecommand \@ifxundefined [1]{%
 \ifx #1\undefined \expandafter \@firstoftwo
 \else \expandafter \@secondoftwo
\fi
}%
\providecommand \@ifnum [1]{%
 \ifnum #1\expandafter \@firstoftwo
 \else \expandafter \@secondoftwo
\fi
}%
\providecommand \enquote [1]{``#1''}%
\providecommand \bibnamefont  [1]{#1}%
\providecommand \bibfnamefont [1]{#1}%
\providecommand \citenamefont [1]{#1}%
\providecommand\href[0]{\@sanitize\@href}%
\providecommand\@href[1]{\endgroup\@@startlink{#1}\endgroup\@@href}%
\providecommand\@@href[1]{#1\@@endlink}%
\providecommand \@sanitize [0]{\begingroup\catcode`\&12\catcode`\#12\relax}%
\@ifxundefined \pdfoutput {\@firstoftwo}{%
 \@ifnum{\z@=\pdfoutput}{\@firstoftwo}{\@secondoftwo}%
}{%
 \providecommand\@@startlink[1]{\leavevmode\special{html:<a href="#1">}}%
 \providecommand\@@endlink[0]{\special{html:</a>}}%
}{%
 \providecommand\@@startlink[1]{%
  \leavevmode
  \pdfstartlink
   attr{/Border[0 0 1 ]/H/I/C[0 1 1]}%
   user{/Subtype/Link/A<</Type/Action/S/URI/URI(#1)>>}%
  \relax
 }%
 \providecommand\@@endlink[0]{\pdfendlink}%
}%
\providecommand \url  [0]{\begingroup\@sanitize \@url }%
\providecommand \@url [1]{\endgroup\@href {#1}{\urlprefix}}%
\providecommand \urlprefix [0]{URL }%
\providecommand \Eprint[0]{\href }%
\@ifxundefined \urlstyle {%
  \providecommand \doi [1]{doi:\discretionary{}{}{}#1}%
}{%
  \providecommand \doi [0]{doi:\discretionary{}{}{}\begingroup
  \urlstyle{rm}\Url }%
}%
\providecommand \doibase [0]{http://dx.doi.org/}%
\providecommand \Doi[1]{\href{\doibase#1}}%
\providecommand \bibAnnote [3]{%
  \BibitemShut{#1}%
  \begin{quotation}\noindent
    \textsc{Key:}\ #2\\\textsc{Annotation:}\ #3%
  \end{quotation}%
}%
\providecommand \bibAnnoteFile [2]{%
  \IfFileExists{#2}{\bibAnnote {#1} {#2} {\input{#2}}}{}%
}%
\providecommand \typeout [0]{\immediate \write \m@ne }%
\providecommand \selectlanguage [0]{\@gobble}%
\providecommand \bibinfo [0]{\@secondoftwo}%
\providecommand \bibfield [0]{\@secondoftwo}%
\providecommand \translation [1]{[#1]}%
\providecommand \BibitemOpen[0]{}%
\providecommand \bibitemStop [0]{}%
\providecommand \bibitemNoStop [0]{.\EOS\space}%
\providecommand \EOS [0]{\spacefactor3000\relax}%
\providecommand \BibitemShut [1]{\csname bibitem#1\endcsname}%
\bibitem{Lee}%
  \BibitemOpen
  \bibfield{author}{%
  \bibinfo {author} {\bibfnamefont{C.}~\bibnamefont{Lee}}, \bibinfo {author}
  {\bibfnamefont{P.}~\bibnamefont{Ghosez}},\ and\ \bibinfo {author}
  {\bibfnamefont{X.}~\bibnamefont{Gonze}},\ }%
  \bibfield{journal}{%
  \Doi{10.1103/PhysRevB.50.13379}{\bibinfo {journal} {Phys. Rev. B}}\ }%
  \textbf{\bibinfo {volume} {50}},\ \bibinfo {pages} {13379} (\bibinfo {year}
  {1994})%
  \bibAnnoteFile{NoStop}{Lee}%
\bibitem{Traylor}%
  \BibitemOpen
  \bibfield{author}{%
  \bibinfo {author} {\bibfnamefont{J.~G.}\ \bibnamefont{Traylor}}, \bibinfo
  {author} {\bibfnamefont{H.~G.}\ \bibnamefont{Smith}}, \bibinfo {author}
  {\bibfnamefont{R.~M.}\ \bibnamefont{Nicklow}},\ and\ \bibinfo {author}
  {\bibfnamefont{M.~K.}\ \bibnamefont{Wilkinson}},\ }%
  \bibfield{journal}{%
  \Doi{10.1103/PhysRevB.3.3457}{\bibinfo {journal} {Phys. Rev. B}}\ }%
  \textbf{\bibinfo {volume} {3}},\ \bibinfo {pages} {3457} (\bibinfo {year}
  {1971})%
  \bibAnnoteFile{NoStop}{Traylor}%
\bibitem{Montanari}%
  \BibitemOpen
  \bibfield{author}{%
  \bibinfo {author} {\bibfnamefont{B.}~\bibnamefont{Montanari}}\ and\ \bibinfo
  {author} {\bibfnamefont{N.~M.}\ \bibnamefont{Harrision}},\ }%
  \bibfield{journal}{%
  \bibinfo {journal} {J. Phys.: Condens. Matter}\ }%
  \textbf{\bibinfo {volume} {16}},\ \bibinfo {pages} {273} (\bibinfo {year}
  {2004})%
  \bibAnnoteFile{NoStop}{Montanari}%
\bibitem{Montanari2}%
  \BibitemOpen
  \bibfield{author}{%
  \bibinfo {author} {\bibfnamefont{B.}~\bibnamefont{Montanari}}\ and\ \bibinfo
  {author} {\bibfnamefont{N.~M.}\ \bibnamefont{Harrision}},\ }%
  \bibfield{journal}{%
  \bibinfo {journal} {Chem. Phys. Lett.}\ }%
  \textbf{\bibinfo {volume} {364}},\ \bibinfo {pages} {528} (\bibinfo {month}
  {January}\ \bibinfo {year} {2002})%
  \bibAnnoteFile{NoStop}{Montanari2}%
\bibitem{Liu}%
  \BibitemOpen
  \bibfield{author}{%
  \bibinfo {author} {\bibfnamefont{Y.}~\bibnamefont{Liu}}, \bibinfo {author}
  {\bibfnamefont{L.}~\bibnamefont{Ni}}, \bibinfo {author}
  {\bibfnamefont{Z.}~\bibnamefont{Ren}}, \bibinfo {author}
  {\bibfnamefont{G.}~\bibnamefont{Xu}}, \bibinfo {author}
  {\bibfnamefont{C.}~\bibnamefont{Song}},\ and\ \bibinfo {author}
  {\bibfnamefont{G.}~\bibnamefont{Han}},\ }%
  \bibfield{journal}{%
  \bibinfo {journal} {J. Phys.: Condens. Matter}\ }%
  \textbf{\bibinfo {volume} {21}},\ \bibinfo {pages} {275901} (\bibinfo {year}
  {2009})%
  \bibAnnoteFile{NoStop}{Liu}%
\bibitem{felich}%
  \BibitemOpen
  \bibfield{author}{%
  \bibinfo {author} {\bibfnamefont{A.}~\bibnamefont{Gr\"unebohm}}, \bibinfo
  {author} {\bibfnamefont{C.}~\bibnamefont{Ederer}},\ and\ \bibinfo {author}
  {\bibfnamefont{P.}~\bibnamefont{Entel}},\ }%
  \bibinfo {note} {unpublished}%
  \bibAnnoteFile{NoStop}{felich}%
\bibitem{Samara}%
  \BibitemOpen
  \bibfield{author}{%
  \bibinfo {author} {\bibfnamefont{G.~A.}\ \bibnamefont{Samara}}\ and\ \bibinfo
  {author} {\bibfnamefont{P.~S.}\ \bibnamefont{Peercy}},\ }%
  \bibfield{journal}{%
  \Doi{10.1103/PhysRevB.7.1131}{\bibinfo {journal} {Phys. Rev. B}}\ }%
  \textbf{\bibinfo {volume} {7}},\ \bibinfo {pages} {1131} (\bibinfo {year}
  {1973})%
  \bibAnnoteFile{NoStop}{Samara}%
\bibitem{Mitev}%
  \BibitemOpen
  \bibfield{author}{%
  \bibinfo {author} {\bibfnamefont{P.~D.}\ \bibnamefont{Mitev}}, \bibinfo
  {author} {\bibfnamefont{K.}~\bibnamefont{Hermansson}}, \bibinfo {author}
  {\bibfnamefont{B.}~\bibnamefont{Montanari}},\ and\ \bibinfo {author}
  {\bibfnamefont{K.}~\bibnamefont{Refson}},\ }%
  \bibfield{journal}{%
  \Doi{10.1103/PhysRevB.81.134303}{\bibinfo {journal} {Phys. Rev. B}}\ }%
  \textbf{\bibinfo {volume} {81}},\ \bibinfo {pages} {134303} (\bibinfo {year}
  {2010})%
  \bibAnnoteFile{NoStop}{Mitev}%
\bibitem{Kresse1}%
  \BibitemOpen
  \bibfield{author}{%
  \bibinfo {author} {\bibfnamefont{G.}~\bibnamefont{Kresse}}\ and\ \bibinfo
  {author} {\bibfnamefont{J.}~\bibnamefont{Furthm\"uller}},\ }%
  \bibfield{journal}{%
  \bibinfo {journal} {Phys. Rev. B}\ }%
  \textbf{\bibinfo {volume} {54}},\ \bibinfo {pages} {11169} (\bibinfo {year}
  {1996})%
  \bibAnnoteFile{NoStop}{Kresse1}%
\bibitem{Blochl}%
  \BibitemOpen
  \bibfield{author}{%
  \bibinfo {author} {\bibfnamefont{P.~E.}\ \bibnamefont{Bl\"ochl}},\ }%
  \bibfield{journal}{%
  \bibinfo {journal} {Phys. Rev. B}\ }%
  \textbf{\bibinfo {volume} {50}},\ \bibinfo {pages} {17953} (\bibinfo {year}
  {1994})%
  \bibAnnoteFile{NoStop}{Blochl}%
\bibitem{PBE}%
  \BibitemOpen
  \bibfield{author}{%
  \bibinfo {author} {\bibfnamefont{J.~P.}\ \bibnamefont{Perdew}}, \bibinfo
  {author} {\bibfnamefont{K.}~\bibnamefont{Burke}},\ and\ \bibinfo {author}
  {\bibfnamefont{M.}~\bibnamefont{Ernzerhof}},\ }%
  \bibfield{journal}{%
  \bibinfo {journal} {Phys. Rev. Lett.}\ }%
  \textbf{\bibinfo {volume} {77}},\ \bibinfo {pages} {3865} (\bibinfo {year}
  {1996})%
  \bibAnnoteFile{NoStop}{PBE}%
\bibitem{bornstoer}%
  \BibitemOpen
  \bibfield{author}{%
  \bibinfo {author}
  {\bibfnamefont{M.}~\bibnamefont{Gajdo\ifmmode~\check{s}\else \v{s}\fi{}}},
  \bibinfo {author} {\bibfnamefont{K.}~\bibnamefont{Hummer}}, \bibinfo {author}
  {\bibfnamefont{G.}~\bibnamefont{Kresse}}, \bibinfo {author}
  {\bibfnamefont{J.}~\bibnamefont{Furthm\"uller}},\ and\ \bibinfo {author}
  {\bibfnamefont{F.}~\bibnamefont{Bechstedt}},\ }%
  \bibfield{journal}{%
  \Doi{10.1103/PhysRevB.73.045112}{\bibinfo {journal} {Phys. Rev. B}}\ }%
  \textbf{\bibinfo {volume} {73}},\ \bibinfo {pages} {045112} (\bibinfo {year}
  {2006})%
  \bibAnnoteFile{NoStop}{bornstoer}%
\bibitem{Burdett}%
  \BibitemOpen
  \bibfield{author}{%
  \bibinfo {author} {\bibfnamefont{J.~K.}\ \bibnamefont{Burdett}}, \bibinfo
  {author} {\bibfnamefont{T.}~\bibnamefont{Hughbanks}},\ and\ \bibinfo {author}
  {\bibfnamefont{J.}~\bibnamefont{Gordon}},\ }%
  \bibfield{journal}{%
  \bibinfo {journal} {J. Am. Chem. Soc}\ }%
  \textbf{\bibinfo {volume} {109}},\ \bibinfo {pages} {3639–3646} (\bibinfo
  {year} {1987})%
  \bibAnnoteFile{NoStop}{Burdett}%
\bibitem{footnote}%
  \BibitemOpen
  \bibinfo {note} {For simplicity, we use the optical $A_{2u}$ eigenmode of the
  unstrained system, even though under (110) strain this mode mixes with two
  $B_{1u}$ modes.}%
  \bibAnnoteFile{Stop}{footnote}%
\bibitem{Antons}%
  \BibitemOpen
  \bibfield{author}{%
  \bibinfo {author} {\bibfnamefont{A.}~\bibnamefont{Antons}}, \bibinfo {author}
  {\bibfnamefont{J.~B.}\ \bibnamefont{Neaton}}, \bibinfo {author}
  {\bibfnamefont{K.~M.}\ \bibnamefont{Rabe}},\ and\ \bibinfo {author}
  {\bibfnamefont{D.}~\bibnamefont{Vanderbilt}},\ }%
  \bibfield{journal}{%
  \Doi{10.1103/PhysRevB.71.024102}{\bibinfo {journal} {Phys. Rev. B}}\ }%
  \textbf{\bibinfo {volume} {71}},\ \bibinfo {pages} {024102} (\bibinfo {year}
  {2005})%
  \bibAnnoteFile{NoStop}{Antons}%
\bibitem{Dieguez/Rabe/Vanderbilt:2005}%
  \BibitemOpen
  \bibfield{author}{%
  \bibinfo {author} {\bibfnamefont{O.}~\bibnamefont{Di{\'{e}}guez}}, \bibinfo
  {author} {\bibfnamefont{K.~M.}\ \bibnamefont{Rabe}},\ and\ \bibinfo {author}
  {\bibfnamefont{D.}~\bibnamefont{Vanderbilt}},\ }%
  \bibfield{journal}{%
  \bibinfo {journal} {Phys. Rev. B}\ }%
  \textbf{\bibinfo {volume} {72}},\ \bibinfo {pages} {144101} (\bibinfo {year}
  {2005})%
  \bibAnnoteFile{NoStop}{Dieguez/Rabe/Vanderbilt:2005}%
\bibitem{Shojaee}%
  \BibitemOpen
  \bibfield{author}{%
  \bibinfo {author} {\bibfnamefont{E.}~\bibnamefont{Shojaee}}\ and\ \bibinfo
  {author} {\bibfnamefont{M.~R.}\ \bibnamefont{Mohammadizadeh}},\ }%
  \bibfield{journal}{%
  \bibinfo {journal} {J. Phys.: Condens. Matter}\ }%
  \textbf{\bibinfo {volume} {22}},\ \bibinfo {pages} {015401} (\bibinfo {year}
  {2009})%
  \bibAnnoteFile{NoStop}{Shojaee}%
\end{thebibliography}%

\end{document}